\documentclass[a4paper,twocolumn,preprintnumbers,citeautoscript,newabstract,aps%,superscriptaddress%,nofootinbib
]{revtex4} %floatfix,floats
\usepackage{amsmath,amsfonts,amssymb,graphics}
\usepackage[normalem]{ulem}
\usepackage{graphicx}
\usepackage{units}
\usepackage{mathrsfs}
\usepackage{bbm}
\usepackage{pifont}
\usepackage{braket}
\usepackage{units}
\usepackage{footmisc}
\usepackage[dvipsnames]{xcolor}
\usepackage{mathtools}
\usepackage{xspace}
\usepackage{booktabs}

\usepackage[bookmarks=false,hyperfootnotes=false]{hyperref}%\usepackage{hyperref}

\definecolor{nicered}{rgb}{0.5,.0,.0}
\definecolor{darkblue}{rgb}{0,.1,.9}
\definecolor{lightblue}{rgb}{0,.1,.6}
 \definecolor{darkgreen}{rgb}{0.0,0.2,0.0}
\hypersetup{colorlinks, citecolor=darkgreen ,linkcolor=darkblue, urlcolor=lightblue}

\definecolor{darkgreen}{rgb}{0.0, 0.6, 0.2}

\allowdisplaybreaks[1]

\begin{document}

%\preprint{preprint#\\}
\title{\textbf{\boldmath \Large \hspace{-1cm}\mbox{GRB 221009A Gamma Rays from Radiative Decay of Heavy Neutrinos?}\unboldmath}}

\author{Alexei Y. Smirnov}
\email[]{smirnov@mpi-hd.mpg.de}
\author{Andreas Trautner}
\email[]{trautner@mpi-hd.mpg.de}
\affiliation{\vspace{0.2cm}Max-Planck-Institut f\"ur Kernphysik, Saupfercheckweg 1, 69117 Heidelberg, Germany}
%
%\date{\today}

\begin{abstract}
We consider a mechanism which allows to decrease attenuation of high energy gamma ray flux from gamma ray burst GRB 221009A. 
The mechanism is based on the existence of a heavy $m_N\sim0.1\,\mathrm{MeV}$ mostly sterile neutrino $N$ which mixes with active neutrinos. 
$N$'s are produced in GRB in $\pi$ and $K$ decays via mixing with $\nu_\mu$. They undergo the radiative decay $N\rightarrow \nu \gamma$ on the way to the Earth.
The usual exponential attenuation of gamma rays is lifted to an attenuation inverse in the optical depth.
Various restrictions on this scenario are discussed. We find that the high energy $\gamma$ events at $18\,\mathrm{TeV}$ and potentially 
$251\,\mathrm{TeV}$ can be explained if (i) the GRB active neutrino fluence is close to the observed limit, 
(ii) the branching ratio of $N\rightarrow \nu \gamma$ is at least of the order 10\%. 
\end{abstract}

\maketitle\widowpenalty10000\clubpenalty10000
\textit{Introduction.---} Recently GRB 221009A set a new record for the brightest gamma ray burst ever detected. 
The initial detection was by BAT, XRT, UVOT on \textit{Swift}, as well as GBM and LAT on \textit{Fermi} satellite, see~\cite{GCN}.
The redshift was determined by X-shooter of VLT (GCN 32648) as well as GTC (GCN 32686) to be~$z=0.1505$ corresponding to a distance of $d\approx645\,\mathrm{Mpc}$.
LHAASO's WCDA as well as KM2A instrument detected $\mathcal{O}(5000)$ photons 
with $E_\gamma\gtrsim500\,\mathrm{GeV}$ from GRB 221009A within $2000\,\mathrm{s}$ 
after the initial outburst (GCN 32677). The photon energies reconstructed by LHAASO extend up to $18\,\mathrm{TeV}$ 
(the relative error of energy determination at $18\,\mathrm{TeV}$ is roughly $40\%$~\cite{Ma:2022aau}), 
and even an observation of a single candidate $\gamma$ ray with an energy of $251\,\mathrm{TeV}$ 
has been reported by Carpet-2 at Baksan Neutrino Observatory~\cite{Dzhappuev:2022}. 

These observations are puzzling because the flux of such high energy $\gamma$ rays should be 
severely attenuated in the intergalactic medium by electron pair 
production on background photons~\cite{Nikishov:1962,Gould:1966pza,Fazio:1970pr}. Standard propagation 
models~\cite{Franceschini:2008tp,Finke:2010,Kneiske:2010,Dominguez:2011,Gilmore:2012,Stecker:2016fsg,Saldana-Lopez:2020qzx} 
typically give optical depths of $\tau\sim5(15)$ for photons of 
$E_\gamma\sim10(18)\,\mathrm{TeV}$, see~\cite{Baktash:2022gnf} and references therein.
This attenuation could be overcome in beyond the Standard Model scenarios 
with axion-photon mixing~\cite{Galanti:2022pbg,Baktash:2022gnf,Lin:2022ocj,Troitsky:2022xso,Nakagawa:2022wwm,Zhang:2022zbm,Gonzalez:2022opy}
(see~\cite{Troitsky:2016akf} for a review) or violation of Lorentz invariance
~\cite{Li:2022wxc,Baktash:2022gnf,Finke:2022swf} (see~\cite{Martinez-Huerta:2020cut} for a review). 
GRB 221009A observations have also triggered further investigations of GRBs as source of UHECR~\cite{Das:2022gon,AlvesBatista:2022kpg,Mirabal:2022ipw},
Earth ionospheric distortions~\cite{Hayes:2022dhp}, and the intergalactic magnetic field~\cite{Xia:2022uua}.

Here, we will consider an entirely different explanation of the observed excess of high energy $\gamma$ rays
based on the existence of a heavy ($\mathcal{O}(0.1)\mathrm{MeV}$ mass scale) mostly sterile neutrino $N$ which mixes with active neutrinos.
Heavy neutrinos are produced in GRB via mixing and then undergo the radiative decay $N\rightarrow\nu\gamma$ on the way to Earth. 
This produces additional high energy flux of $\gamma$ rays that would experience less attenuation.

\smallskip
\textit{Fluxes of $\nu$ and $N$.---} 
GRBs are powerful sources of high energy neutrinos~\cite{Waxman:1997ti}.
However, the predicted neutrino fluxes $\Phi_\nu$ are highly uncertain, see e.g.~\cite{He:2012tq,Denton:2017jwk} 
and~\cite{Kimura:2022zyg} for a review, with a conservative uncertainty estimate of larger than two orders of magnitude.
The time integrated fluxes (fluences) could reach $E_\nu^2\Phi_\nu^{\mathrm{int}}\simeq\mathcal{O}(10^{-5})\,\mathrm{TeV}\mathrm{cm}^{-2}$ 
at energies of $\mathcal{O}(\mathrm{TeV})$ and the general expectation is that $E_{\nu}^2\Phi_\nu^{\mathrm{int}}$ is rising 
for energies up to $\mathcal{O}(10^{3})\,\mathrm{TeV}$. 

An upper bound on the neutrino fluence of GRB 221009A has been set from the non-observation of track-like neutrino 
events in the energy range $0.8\,\mathrm{TeV}\div1\,\mathrm{PeV}$ by \textit{IceCube} and is given by~(GCN 32665 and~\cite{Ai:2022kvd,Murase:2022vqf})
\begin{equation}\label{eq:nufluence}
E_{\nu}^2\,\Phi_\nu^{\mathrm{int}}<3.9\times10^{-5}\,\mathrm{TeV}\mathrm{cm}^{-2}\;.
\end{equation}

Let us introduce the ratio of the neutrino flux $\Phi_\nu$ to the unattenuated $\gamma$ flux $\Phi^0_\gamma$,
\begin{equation}
r_{\nu\gamma}\equiv\frac{\Phi_\nu}{\Phi^0_\gamma}\;. 
\end{equation} 
The unattenuated $\gamma$ flux of GRB 221009A can be obtained by extrapolating the flux measured by \textit{Fermi}-LAT~(GCN 32658)
in the energy range $(0.1\div1)\mathrm{GeV}$ to higher ($\mathrm{TeV}$ scale) energies~\cite{Baktash:2022gnf}:
\begin{equation}\label{eq:immediate_flux}
 \Phi^0_\gamma(E_\gamma) = \frac{2.1\times10^{-6}}{\mathrm{cm}^{2}\mathrm{s}\,\mathrm{TeV}}\left(\frac{E_\gamma}{\mathrm{TeV}}\right)^{-1.87\pm0.04}\;. 
\end{equation}

Dividing the \textit{IceCube} bound on neutrino fluence~\eqref{eq:nufluence} by the $\Delta t\simeq600\,\mathrm{s}$ long period of most 
intense $\gamma$ emission we obtain an average neutrino flux $\Phi_\nu=\Phi^{\mathrm{int}}_\nu/\Delta t$ and consequently
a flux ratio of $r_{\nu\gamma}\lesssim3\times10^{-2}$. For shorter periods of time much larger flux ratios are possible.
Notice that the total number of events is given by the integral over time and, therefore, does not depend on the value of $\Delta t$.

Since GRB neutrinos are predominantly produced in pion and muon decays~\cite{Kimura:2022zyg} the flux of heavy neutrinos for $m_N \lesssim 1\,\mathrm{MeV}$
can be parameterized as 
\begin{equation}
r_{N\nu}\equiv\frac{\Phi_N}{\Phi_\nu}=\frac{\sum_{\ell = e, \mu} |U_{N\ell}|^2 \Phi_{\nu_\ell}}{\sum_{\ell = e, \mu}\Phi_{\nu_\ell}}\;.
\end{equation}
If $N$ would exclusively mix with $\nu_\mu$ and the total highest energy neutrino flux is dominated by $\nu_\mu$, 
then $r_{N\nu}=|U_{N\mu}|^2$ is simply given by the corresponding mixing matrix element. We adopt this case as a benchmark. 

The angular dispersion of $\gamma$'s produced in $N$ decay is $\Theta\simeq m_N/E_N\sim10^{-8}$ with energy $E_N$. 
If the GRB jet opening angle is bigger than $\Theta$, then there is no additional suppression of the $\gamma$ flux from $N$ at the Earth.

\smallskip
\textit{Propagation scenario.---}
Let us compute the $\gamma$ flux at Earth originating from $N$ decays.
In terms of the total decay rate $\Gamma_N$ %, as well its relative velocity $\beta\leq1$,
the decay length is given by
\begin{equation}
 \lambda_N = \frac{E_N}{\Gamma_N\,m_N}\;. %\beta
\end{equation}
The probability that an individual $N$ decays in the distance interval $[x,x + \mathrm{d}x]$
and the produced photon reaches the Earth equals
\begin{equation}\label{eq:ind}
B_\gamma\,\mathrm{e}^{- x/\lambda_N}\,\frac{\mathrm{d}x}{\lambda_N}\,\mathrm{e}^{-(d-x)/\lambda_\gamma}\;,
\end{equation}
where $B_\gamma$ is the branching ratio of radiative decay,
and the last factor describes the survival probability of $\gamma$ in terms of its absorption length~$\tau\equiv d/\lambda_\gamma$.
Multiplying the expression in eq.~\eqref{eq:ind}
by the $N$ flux $\Phi_N$ and integrating over $x$, we find the $N$-induced $\gamma$ flux
\begin{equation}\label{eq:ind2}
\Phi^{(N)}_\gamma = \Phi_N B_\gamma \frac{1}{\lambda_N/\lambda_\gamma -1}
\left[\mathrm{e}^{-d/\lambda_N} - \mathrm{e}^{-d/\lambda_\gamma} \right].
\end{equation}
Normalizing~\eqref{eq:ind2} to $\Phi^0_\gamma$, the direct unattenuated $\gamma$ flux, we find
\begin{equation}\label{eq:ind_exact}
\frac{\Phi^{(N)}_\gamma}{\Phi^0_\gamma} = B_\gamma\,\frac{\Phi_N}{\Phi^0_\gamma}\,
\frac{1}{\tau \lambda_N/d -1}
\left[\mathrm{e}^{-d/\lambda_N} - \mathrm{e}^{- \tau} \right]\;.
\end{equation}
Varying $d/\lambda_N$ we find that the maximal flux is obtained for $d/\lambda_N \approx 1$.
Using this and the flux ratios $r_{N\nu}$ and $r_{\nu\gamma}$ defined earlier,
as well as expanding in $\tau\gg1$ as expected for high energy $\gamma$ rays we obtain
\begin{equation}\label{eq:flux_extimate}
\frac{\Phi^{(N)}_\gamma}{\Phi^0_\gamma} \;\approx \;B_\gamma\;r_{N\nu}\; r_{\nu\gamma}\; \frac{0.37}{\tau}\;.
\end{equation}
Recall that the $\gamma$ flux produced directly in GRB is attenuated as
\begin{equation}\label{eq:fdirect}
\frac{\Phi_\gamma^d}{\Phi_\gamma^0} = \mathrm{e}^{-d/\lambda_\gamma}  = \mathrm{e}^{-\tau}\;.
\end{equation}
Eq.~\eqref{eq:flux_extimate} and \eqref{eq:fdirect} clearly show how the usual damping of the high energy $\gamma$ ray flux, exponential in $\tau$, can be overcome
by the presence of decaying heavy neutrinos. 

Let us underline that there is strong energy dependence in all of these expressions: $\Phi_{N}$ and $\lambda_N$ 
depend on $E_N$, while $\lambda_\gamma$ (or equivalently $\tau$) strongly depends on $E_\gamma$.
The explicit $E_N$ dependence of the attenuation factor can be displayed writing $\lambda_N/d=E_N/E_N^d$ with $E_N^d\equiv\Gamma_N m_N d$ being the energy at which $\lambda_N=d$.
Then eq.~\eqref{eq:ind_exact}, neglecting the last term in brackets, reduces to
\begin{equation}
 \frac{\Phi^{(N)}_\gamma}{\Phi^0_\gamma} = B_\gamma\,\frac{\Phi_N}{\Phi^0_\gamma}\,
\frac{\mathrm{e}^{-E_N^d/E_N}}{\tau E_N/E_N^d -1}\;.
\end{equation}

\begin{figure}[t]
\centering\includegraphics[width=1\linewidth]{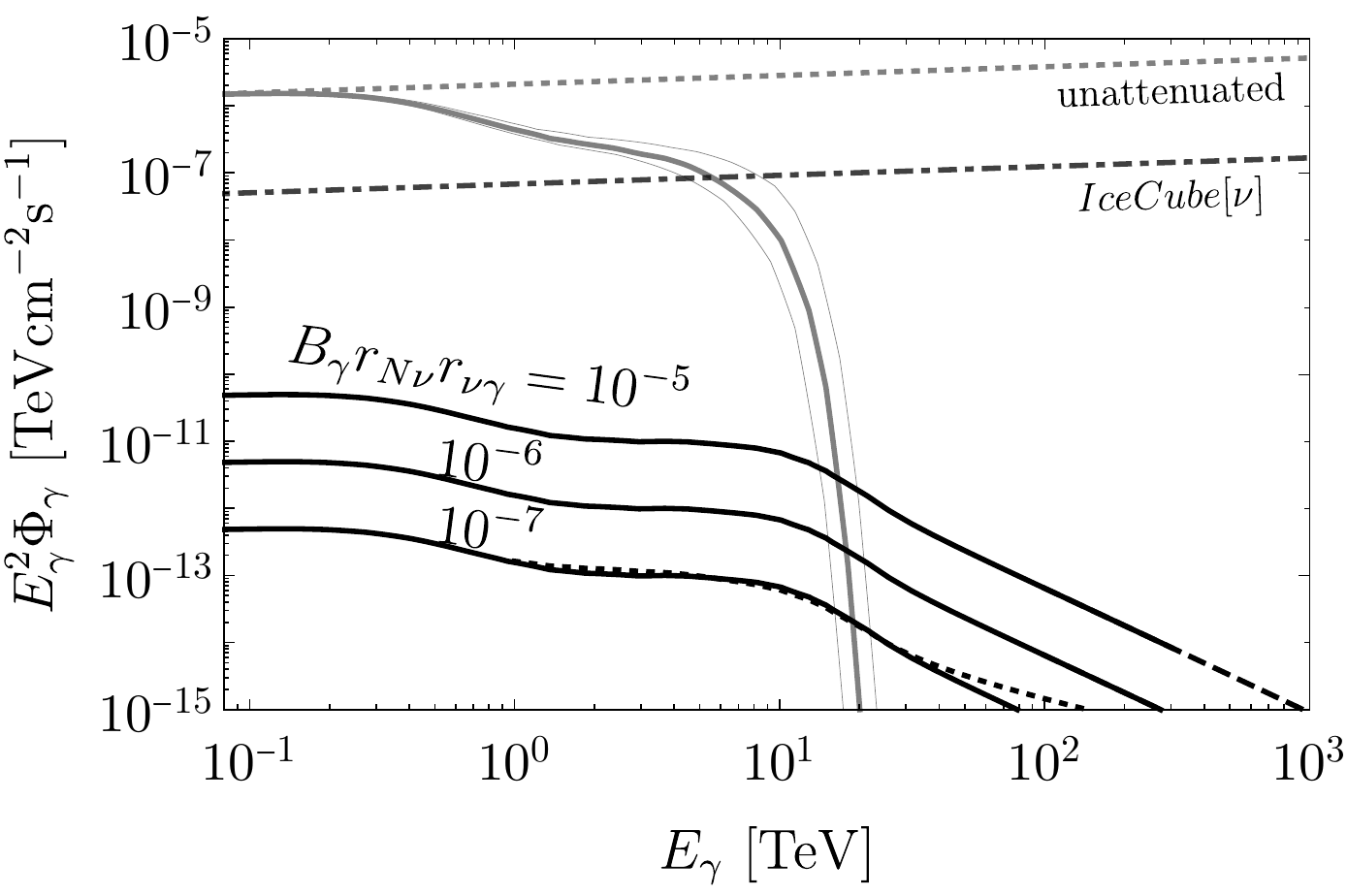}
\caption{\label{fig:Nflux}
The $\gamma$ fluxes from GRB 221009A at the Earth as functions of $E_\gamma$.
Solid black: $\gamma$ flux induced by $N\rightarrow\nu\gamma$ decay for different values of prefactors in 
eq.~\eqref{eq:ind_exact} and~\eqref{eq:flux_extimate} under the assumption that $\lambda_N=d$ at $E_N=40\,\mathrm{TeV}$. 
Gray: direct $\gamma$ flux with uncertainties as obtained with $\tau(E_\gamma)$ from~\cite{Dominguez:2011}.
Dashed gray: unattenuated $\gamma$ flux.
Shown in dash-dotted is also the upper bound on the neutrino flux obtained from the \textit{IceCube}
bound on the neutrino fluence divided by $\Delta t=600\,\mathrm{s}$.
The dashed black line shows the approximation of eq.~\eqref{eq:flux_extimate} for the 
case $B_\gamma r_{N\nu} r_{\nu\gamma}=10^{-7}$.}
\end{figure}

In Fig.~\ref{fig:Nflux} we show the secondary $\gamma$ flux from $N$ decay for GRB 221009A. 
We use the approximation $E_\gamma\approx 0.5\,E_N$, the maximal $\nu$ flux 
allowed by \textit{IceCube}, and the full energy dependence of $\tau(E_\gamma)$ 
as extracted from~\cite{Dominguez:2011} as well as the assumption 
that $\lambda_N=d$ at $E_N=40\,\mathrm{TeV}$.

\smallskip
\textit{Model independent constraints.---}
The radiative decay of heavy neutrinos produces $\gamma$ rays with energy $E_\gamma\lesssim E_N$. 
The existence of non-zero mass of $N$ leads to dispersion of the $\gamma$ signal in time.
Requiring  that $\gamma$'s of highest energies $18\,\mathrm{TeV}$
arrive at the detector within $\Delta t\leq2000\,\mathrm{s}$ the heavy neutrino mass is bounded by
\begin{equation}\label{eq:constraint1}
 m_N\lesssim 4.5\,\mathrm{MeV}\;
 \left(\frac{\Delta t}{2000\,\mathrm{s}}\right)^{\frac12}
 \left(\frac{E_N}{18\,\mathrm{TeV}}\right)\,.
\end{equation}
Note that the detected $\gamma$ rays originating from the lowest energy heavy neutrinos set the most stringent bound here. 
If $\gamma$'s with an energy as low as $0.5\,\mathrm{TeV}$ could be identified to originate from $N$ decay 
this would tighten the bound to $m_N\lesssim0.25\,\mathrm{MeV}$ but such an identification is unlikely 
given the large background from conventional $\gamma$'s in this region. 
Conversely, if a very high energy $\gamma$ is identified with a long time delay event this would hint at a higher $m_N$.
The bound can also be affected by finite interval of pion production and dependence of energy of 
accelerated protons, and therefore pions, on time. Detailed information on arrival time of
$\gamma$'s of different energies will allow to refine the bound.

Requiring $\lambda_N\sim d$ such that a substantial number of decays happen before the heavy neutrinos reach the Earth reads
\begin{equation}\label{eq:constraint2}
 \Gamma_N\,m_N \gtrsim 2\times10^{-31}\,\mathrm{MeV}^2
 \left(\frac{E_N}{18\,\mathrm{TeV}}\right)\,.
\end{equation}

For masses between $10\,\mathrm{keV}$ and a few $\mathrm{MeV}$ there are strong bounds on heavy \textit{active} neutrino radiative decays from SN1987A~\cite{Kolb:1988pe, Oberauer:1993yr}
\begin{equation}\label{eq:SNbound}
 \Gamma_\nu\, B_{\gamma} \lesssim 5\times 10^{-14}\left(\frac{m_\nu}{\mathrm{MeV}}\right)\mathrm{s}^{-1}\;.
\end{equation}
The flux of heavy neutrinos produced by SN1987A can be parametrized by the ratio 
\begin{equation}
 r^{(SN)}_{N\nu}\equiv\frac{\Phi^{(SN)}_N}{\Phi_\nu}\;.
\end{equation}
Naively scaling the limit \eqref{eq:SNbound} by this ratio we obtain the constraint 
\begin{equation}\label{eq:constraint3}
 \frac{\Gamma_N}{m_N} \lesssim \frac{3\times10^{-35}}{ B_{\gamma}\,r^{(SN)}_{N\nu}}\;.
\end{equation}
Combining this with condition~\eqref{eq:constraint2} requires
\begin{equation}\label{eq:BR_bound}
  B_{\gamma}\, r^{(SN)}_{N\nu} \lesssim 1.7\times10^{-4}\left(\frac{m_N}{\mathrm{MeV}}\right)^2\;.
\end{equation}
This shows that saturation of $B_{\gamma}\leq1$ is not excluded by the model independent 
constraints if $r^{(SN)}_{N\nu}\ll r_{N\nu}\approx|U_{N\mu}|^2$, 
which can be the case due to different production mechanisms and flavor composition.

\smallskip
\textit{Model dependent considerations.---}
Due to the mixing with active neutrinos, in the most minimal scenarios $N$ decays via three-body or 
two-body radiative channels with rates, see e.g.~\cite{Zatsepin:1978iy},
\begin{align}
\Gamma_N^{(3)} &\approx \frac{G_\mathrm{F}^2\,m_N^5}{64\,\pi^3}\left|U_{N\mu}\right|^2 \;,&\\ \label{eq:gamma2}
\Gamma_N^{(2)} &\approx \frac{9\,\alpha\,G_\mathrm{F}^2\,m_N^5}{512\,\pi^4}\left|U_{N\mu}\right|^2 \;.&
\end{align}
At face value this gives rise to a branching fraction
\begin{equation}
 B_\gamma\approx \frac98 \frac{\alpha}{\pi} \approx 2.6\times10^{-3}\;.
\end{equation}
Furthermore, one can use the explicit decay rates together with~\eqref{eq:constraint2} in order to obtain
\begin{equation}\label{eq:constraint4}
m_N \gtrsim\frac{0.125\,\mathrm{MeV}}{\left|U_{N\mu}\right|^{\frac13}}\left(\frac{E_N}{18\,\mathrm{TeV}}\right)^{\frac16}\left(\frac{\lambda_N}{d}\right)^{\frac16}\;.
\end{equation}
Note that in case the radiative decay dominates, this result would only mildly tighten to $m_N\gtrsim0.34\,\mathrm{MeV}$ 
due to the strong, sixth order dependence of~\eqref{eq:constraint2} on $m_N$.
The bounds in eq.~\eqref{eq:constraint1} and~\eqref{eq:constraint4} leave a rather narrow range $0.2\lesssim m_N\lesssim4\,\mathrm{MeV}$.

There are strong constraints on the $|U_{N\ell}|^2-m_N$ parameter space derived from energy loss of SN1987A~\cite{Zhou:2015jha,Drewes:2016upu}.
These constraints are subject to theoretical, supernova modelling and observational uncertainties and have recently been subject to further scrutiny~\cite{Syvolap:2019dat,Suliga:2020vpz,Suliga:2021hek} with the conclusion that they are generally not robust~\cite[Sec.~7.1.3]{Abdullahi:2022jlv}.
For large mixing parameter $|U_{N \mu}|^2 \sim 10^{-2} $ a protoneutron star is not transparent to $N$ and so the cooling arguments may not apply for large mixing. 

On the other hand, with such a large mixing $N$'s thermalize in the Early Universe, thus giving $\Delta N_{\mathrm{eff}} \approx 1$ in the epoch of big bang nucleosynthesis~(BBN) thereby changing the ratio of light element abundances, see e.g.~\cite{Boyarsky:2009ix,Fields:2019pfx}.
The BBN bounds can be avoided in specific models with late phase transitions~\cite{Fuller:1990mq,Chacko:2004cz,Mohapatra:2004uy,Vecchi:2016lty}
or by invoking neutral lepton asymmetries~\cite{Foot:1995bm}. 
Another solution arises if the mass of $N$ is above a few $\mathrm{MeV}$ such that it turns non-relativistic 
before BBN.

The life time of $N$ at rest is $10^{2}\div10^{3}$ years which is much
shorter than the time of recombination epoch in the Universe $t_{\mathrm{rec}} = 3\times10^{5}$ years.
Therefore no substantial distortion of the cosmic microwave background is expected.

For this analysis, if we put aside the model dependent cosmological bounds on $|U_{N\mu}|^2$, 
the strongest constraints arise from PMNS unitarity. We adopt as a benchmark $|U_{N\mu}|^2\approx10^{-3}$, see e.g.~\cite{Parke:2015goa}.

The transition magnetic moment can be estimated as
\begin{equation}\label{eq:magneticmoment}
\mu_N \simeq \sqrt{8\pi B_\gamma\Gamma_N/m_N^3}\;.
\end{equation}
The decay rate of $N$ used here for $m_N = 0.2\,\mathrm{MeV}$ and $|U_{N\mu}|^2 =
10^{-3}$ corresponds to the transition magnetic moment $\mu_N \simeq 10^{-15} \mu_{\mathrm{B}}$, where $\mu_{\mathrm{B}}$
is the Bohr magneton. Therefore the strongest bounds on neutrino magnetic moments are satisfied~\cite{Viaux:2013lha,Borexino:2017fbd}.

\smallskip
\textit{Estimation of number of events.---}
In the following we formulate the requirements on the heavy neutrino scenario in order to explain the observed GRB 221009A high energy events. 
The number of events corresponding to the unattenuated $\gamma$ flux $\Phi_\gamma^0$ is directly computed from~\eqref{eq:immediate_flux}.
For an effective area of $1\,\mathrm{km}^2$~\cite{Cui:2014bda,Ma:2022aau} and observation time $2000\,\mathrm{s}$ 
there are approximately $5\times10^{6}$ events in the energy range $(10\div40)\mathrm{TeV}$.

The corresponding flux of $N$-induced $\gamma$ events can be estimated via~\eqref{eq:flux_extimate}.
Using $r_{N\nu}\approx|U_{N\mu}|^2\approx10^{-3}$, $r_{\nu\gamma}\approx10^{-2}$, $\tau\approx10$
and $B_\gamma\approx10^{-3}$ we obtain an expected number of events of $10^{-3}$ in agreement 
with the result of an exact integration using~\eqref{eq:ind_exact} and taking into account the energy 
dependence of $\lambda_N$ and $\tau$. While a detection would nevertheless be unlikely, this still 
corresponds to an increase in the expected number of events by a few orders of magnitude as compared to 
most standard propagation models, cf.~\cite{Baktash:2022gnf}. Note that $\Phi^{(N)}_\gamma$ is only 
linearly suppressed in $\tau$. Hence, the expected number of events at higher energies is tremendously 
increased over standard propagation models. 
For example, for parameters $B_\gamma\,r_{N\nu}\,r_{\nu\gamma}\approx10^{-5}$ we find that the expected 
number of events in the energy range $(40\div500)\mathrm{TeV}$ is $\sim10^{-4}$ while in standard propagation 
models it is suppressed by a factor smaller than $\mathrm{e}^{-80}$.

\smallskip
\textit{Large $B_\gamma$.---}
Note that the expected number of events in the region $(10\div40)\mathrm{TeV}$ can be pushed to $0.1\div1$ if $B_\gamma\approx0.1\div1$. 
This would also increase the number of expected events in $(40\div500)\mathrm{TeV}$ to $\mathcal{O}(10^{-2})$, potentially 
explaining both, LHAASO and Baksan observation. However, misidentification of a galactic foreground still is the more
likely explanation for the $251\,\mathrm{TeV}$ event~\cite{Fraija:2022}.

Large branching ratios for radiative decay of $N$ can be obtained in specific models.
In the left-right symmetric models with right handed current interactions of $N$ the radiative decay rate can be much bigger than that in eq.~\eqref{eq:gamma2}.
In this case the enhancement factor $32 \sin^2 2\xi (m_\mu/m_N)^2$ appears, where $\xi$ is the mixing angle of $W_L$ and $W_R$. 
Taking $\sin^2 2\xi = 2\times10^{-6}$ we obtain the factor $16$. Bigger enhancement ($\sim 100$) can be obtained for tau-lepton mass (which implies $N$ mixing with $\nu_\tau$), 
thus leading to $B_\gamma \simeq 0.2$. Even bigger enhancement can be obtained in the models with charged
scalars (the Zee type models~\cite{Zee:1980ai}, see~\cite{Babu:2020ivd} and references therein for a recent discussion), so that $B_\gamma \simeq 1$ can be achieved.
Larger branching ratios of $N\rightarrow\nu\gamma$ in these models correspond to larger transition magnetic moments
and the corresponding decay widths can be computed with~\eqref{eq:magneticmoment}.
Using \eqref{eq:constraint2} and the strongest laboratory contraints $\mu_N\lesssim3\times10^{-11}\mu_{\mathrm{B}}$~\cite{Borexino:2017fbd}
the lower limit on the mass~\eqref{eq:constraint4} can be relaxed to $m_N\gtrsim10^{-2}\,\mathrm{MeV}$. If the 
strongest astrophysical constraints $\mu_N\leq4.5\times10^{-12}$\cite{Viaux:2013lha} are used (applicable only for $m_N\lesssim20\,\mathrm{keV}$) the 
limit on $m_N$ quantitatively agrees with eq.~\eqref{eq:constraint4}.

For large $B_\gamma$ to be in agreement with the SN1987A constraints, eq.~\eqref{eq:BR_bound} requires $r^{(SN)}_{N\nu}$ to be at least an order of 
magnitude smaller than $r_{N\nu}$, necessitating suppressed heavy neutrino production in supernovae as compared to GRBs.

\smallskip
\textit{Conclusion.---} 
We have considered the production of heavy neutrinos in GRB and their sequential radiative decay
on the way to the Earth. We showed that in this way one can avoid the exponential suppression of the $\gamma$ flux 
with optical depth $\mathrm{e}^{-\tau}$ and obtain $1/\tau$ suppression instead. This give rise to an observable number of highest energy 
events at LHAASO if the mixing angle is large $|U_{N\mu}|^2\sim 10^{-3}$ and branching ratio $B_\gamma \sim (0.1\div1)$.
We find that the mass of $N$ should be in a narrow range $(0.2\div4)\mathrm{MeV}$.

We have discussed constraints on the mixing and branching fractions and find that they are possible 
to meet in specific models. The required value of $B_\gamma$ can be obtained in further extensions
of the Standard Model beyond just mixing of $N$ with $\nu_\mu$.

More refined estimates of the event rate, $\gamma$ spectrum and the available
parameter space are possible by assuming specific forms for the spectral and time dependences of the $\gamma$ fluxes and we plan to return
to this in the future. The publication of a detailed spectrum of the high energy events by LHAASO and
additional observations of future GRB's could clarify the situation.
If the hint for unexplained high energy gamma rays persist the heavy neutrino with the characteristics described here 
could become a worthwhile target for searches in terrestrial laboratories.

\medskip
\begin{acknowledgments}%\vspace{-0.17in}
We thank Evgeny Akhmedov and Sudip Jana for useful conversations.

\medskip
\textbf{\textit{Note added.---}}%
During the completion of this work, ref.~\cite{Cheung:2022luv} appeared on the arXiv which also considers radiative decay of heavy neutrino
as a way to explain GRB 221009A observations. In~\cite{Cheung:2022luv}, $N$ is produced via the transition magnetic moment (and not mixing). 
This leads to a suppression of the $N$ flux by factor $\mu_N m_\pi \sim 10^{-7}$ for the magnetic moment $\mu_N = 3\times10^{-9}\,\mu_{\mathrm{B}}$.
This suppression is too strong to lead to any number of high energy events at LHAASO. Furthermore, such a large value of $\mu_N$ is excluded by 
laboratory and especially astrophysical observations.
\end{acknowledgments}

\bibliographystyle{utphys}
\bibliography{Orbifold}

\end{document}